# OPTICS RECONSTRUCTION IN THE SCL SECTION OF THE FNAL LINAC *

J.-P. Carneiro†, E. Chen, A. Pathak, R. Sharankova, A. Shemyakin
Fermi National Accelerator Laboratory, Batavia, USA

*Abstract*

The Side-Coupled Linac (SCL) section of the FNAL linac accelerates the beam from 117 MeV to 401.5 MeV, operating at 22-24 mA beam current. Transverse focusing is performed by 32 quadrupoles, and the beam orbit is guided by 19 dipole correctors and measured by 29 BPMs. The bunch length is measured in a single location by a Bunch Shape Monitor (BSM). This paper presents a three-step reconstruction of the machine optics. First, the transverse and longitudinal Twiss parameters at the start of the SCL section are determined using quadrupole scans and BSM measurements at different settings of an upstream cavity. Second, the quadrupole calibrations are adjusted based on differential-trajectory measurements. Finally, the beam is propagated along the SCL linac using the code TraceWin. A comparison between TraceWin simulations and the beam envelope measured by the 12 wire scanners of the SCL linac was performed. Transverse and longitudinal beam parameters at the entrance of the transition section will be reported.

## INTRODUCTION

The front-end of the Fermilab linac consists of an Ion Source producing 35 keV $H^-$ beam and a Low Energy Beam Transport (LEBT) that matches the $H^-$ beam produced by the source into a 4-rod RFQ operating at 201.25 MHz. The beam exits the RFQ with an energy of 750 keV and is matched into a Drift Tube Linac (DTL) by a Medium Energy Beam Transport (MEBT) made of two quadrupole doublets and one buncher. The DTL section operates at 201.25 MHz and consists of 207 drift tubes distributed over 5 DTL tanks. At the exit of the DTL Tank5, the $H^-$ beam has an energy of approximately 117 MeV. A 4-meter Transition section made of four quadrupoles and two non-accelerating 805 MHz cavities (the Buncher and the smaller Vernier) insures proper longitudinal matching between the DTL Tank5 and the Side Coupled Linac (SCL) section, which further accelerates the beam to approximately 401.5 MeV. The SCL section consists of seven 805 MHz modules, with each module containing four Side-Coupled Cavities (SCC). Transverse focusing on the DTL, transition and SCL sections is made by quadrupoles.

Upon exiting the SCL section, the beam travels about 70 meters for injection into the 8 GeV Booster synchrotron. During daily operation, the linac outputs 22-24 mA beam in 35 µs long pulses at 15 Hz repetition rates.

In this paper, we present an attempt to reconstruct the Twiss parameters at the start of the Transition section, based on beam measurements and the beam dynamics code TraceWin [1]. In the second part of the paper, we describe a method based on differential trajectory method and the code Linac_Gen [2] for calibration of the Transition and SCL sections quadrupoles. We present our status between measured and reconstructed beam envelopes along the Transition and SCL sections.

## TWISS RECONSTRUCTION

Figure 1 shows a schematic of the Transition section and the start of the SCL section (Module 1, Cavities 1 and 2). We attempted to reconstruct the Twiss at the Beam Position Monitor downstream of the DTL Tank5 (as shown in Figure 1), for normal operation at 23.7 mA. A summary of the reconstructed Twiss parameters at this location is reported in Table 1.

Table 1: Reconstructed beam parameters at the start of the matching section (BPM downstream Tank5, see Figure 1)

| Beam Parameter | Value |
|---|---|
| Energy | 117 MeV |
| Beam current | 23.7 mA |
| Alpha X | -0.49 |
| Beta X | 9.03 mm/mrad |
| Emittance rms norm. X | 0.9 mm-mrad |
| Alpha Y | 0.35 |
| Beta Y | 1.89 mm/mrad |
| Emittance rms norm. Y | 0.8 mm-mrad |
| Alpha Z | -0.5 |
| Beta Z | 8.0 mm/mrad |
| Emittance rms norm. Z | 1.0 mm-mrad |
| Emittance rms norm. Z | 0.45 deg-MeV |

We first started the Twiss reconstruction by performing a calibration of the Buncher and the Vernier during nominal operation. We recorded the signals of the downstream BPMs as a function of the Buncher and Vernier phases. The BPM signals (recorded at a base frequency of 402.5 MHz) were then reconstructed with TraceWin which allowed us to estimate the operating effective gap voltages and synchronous phases of the Buncher and Vernier. We estimate from these BPM studies that the Buncher and Vernier are operating daily with an effective gap voltage of respectively $E_0T = 1.68$ MV/m and $E_0T = 2.95$ MV/m and that both cavities operate with a synchronous phase of -90 deg.

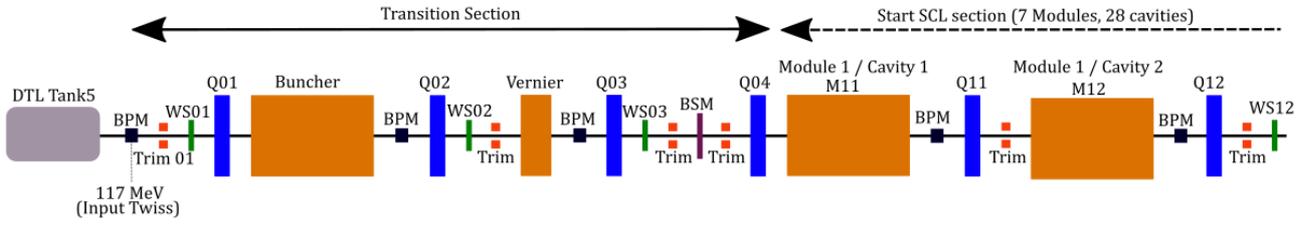

Figure 1: Schematic of the FNAL Transition section (117 MeV) and start of the SCL section.

Once the Buncher and Vernier calibrated, we performed a longitudinal emittance estimate by measuring the rms bunch length at the Bunch Shape Monitor (BSM, located between Q03 and Q04 in Fig. 1) as a function of the Buncher phase (with the Vernier off). This longitudinal phase scan was then reconstructed with TraceWin taking longitudinal input Twiss (at the BPM downstream of the DTL Tank5) of $\alpha_z = -0.5$, $\beta_z = 8$ mm/mrad and a rms normalized longitudinal emittance of 1.0 mm-mrad (or 0.45 deg-MeV). A description of the BSM measurements is reported in Ref. [3].

To reconstruct the transverse Twiss, we performed a quadrupole scan using the last quadrupole of the Transition section (Q04, Fig.1) and the first Wire Scanner of the SCL section (WS12, Fig. 1). Figure 2 shows the Horizontal (a) and Vertical (b) quadrupole scans Q04-WS12.

During the quadrupole scans, the Buncher and Vernier were operating at their nominal values (as previously described) and the quadrupoles Q01, Q02 and Q03 were optimized to enable a waist in both the Horizontal plane and Vertical plane at WS12, during the scan of Q04. All the elements between Q04 and WS12 were turned off, making the distance between Q04 and WS12 (~3348 mm) a drift. We extracted the Twiss at Q04 from the usual quadratic fits (without space charge) of the rms beam sizes at the WS12. These Twiss were then backpropagated to the initial point (BPM downstream Tank 5) with TraceWin, without space charge. The initial Twiss were then adjusted manually in TraceWin to mimic the measured rms sizes at W12 as a function of Q04, in forward propagation and in the presence of space charge (23.7 mA). The horizontal Twiss at the initial point were determined to be: $\alpha_x = -0.49$, $\beta_x = 9.03$ mm/mrad and a rms normalized horizontal emittance of 0.9 mm-mrad. The vertical Twiss at the initial point were determined to be: $\alpha_y = -0.35$, $\beta_y = 1.89$ mm/mrad and a rms normalized vertical emittance of 0.8 mm-mrad. These Twiss are reported in Table 1 and Figure 2.

## ENVELOPE IN TRANSITION SECTION

Figure 3 shows the measured envelope in the Transition section at the Wire Scanners WS01, W02 and WS03, as shown in Figure 1. This envelope measurement was performed the same day as the above mentioned horizontal and vertical quad scans. This envelope corresponds to the daily operation of the linac, with quadrupole settings in the

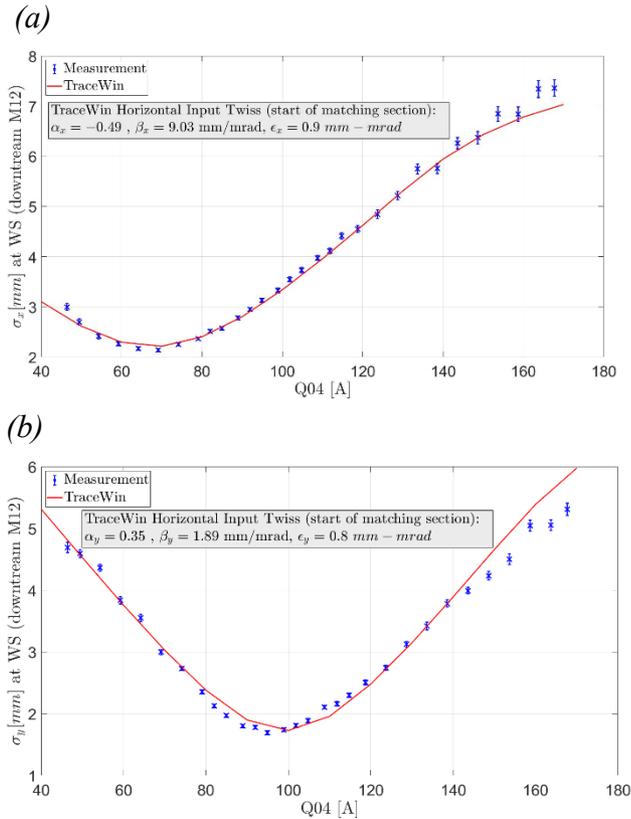

Figure 2: (a) Horizontal and (b) Vertical quadrupole scans Q04-WS12 with TraceWin reconstruction.

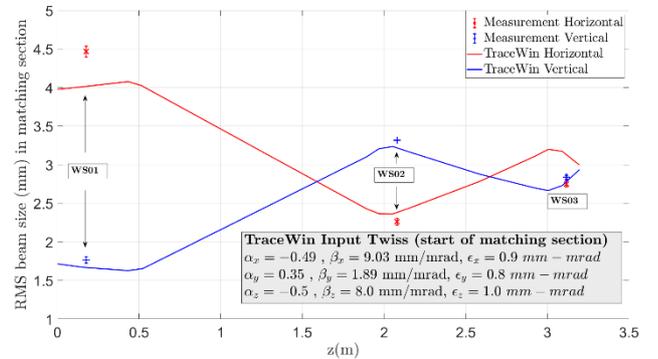

Figure 3: Measured envelope at the 3 Wire Scanners of the matching section and comparison with TraceWin.

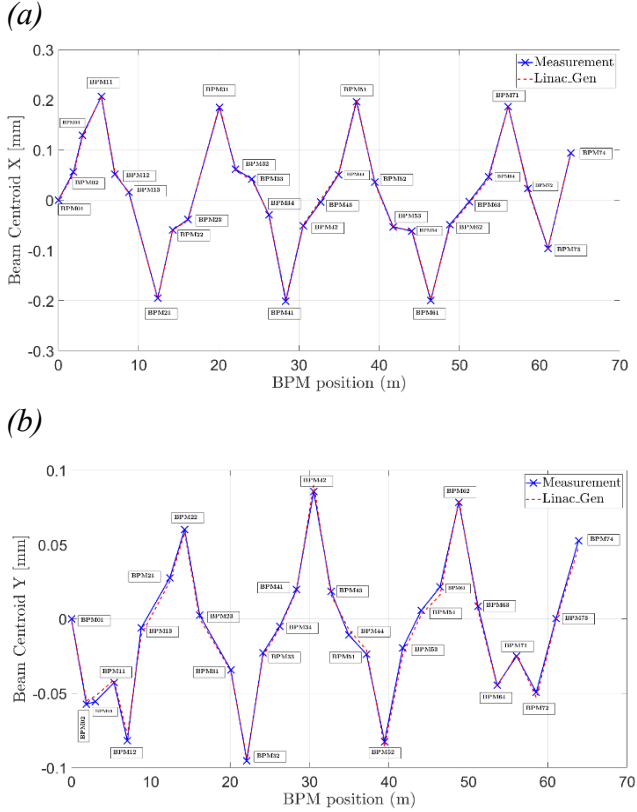

Figure 4: (a) Horizontal and (b) Vertical beam centroid motion measured and reconstructed with Linac_Gen, at all 29 BPMs of the Transition and SCL sections, for a current of 0.1 A in Trim01 (downstream DTL Tank5, see Fig1).

Transition section different than those set for the quadrupole scans. We can observe in Figure 3 that the measured and simulated envelope is in good agreement in the vertical plane, but the horizontal plane shows 10% disagreement in WS01 and 13% in WS03. We think that the calibration of the quadrupole in the Transition section may explain this disagreement. A method for quadruple calibration described below is presently under development at Fermilab.

## QUADRUPOLES CALIBRATION

We are presently developing a tool at Fermilab to perform beam-based calibration of quadrupoles through the differential trajectory method. This method consists of measuring a reference trajectory and using a steering magnet to deviate the beam and obtain a differential trajectory at each BPM. The Fermilab code Linac_Gen is then used to adjust the gradient of each quadrupole that matches the measured beam centroid. This beam-based calibration of quadrupoles is expected to be used during the commissioning of the PIP-II linac [4].

Figure 4 shows the horizontal and vertical measured beam centroid motion at each one of the 29 BPMs (spread from downstream of Tank 5 (BPM01) to downstream of the SCL Cavity 4 of Module 7 (BPM74)), in the case of (a) a horizontal beam deviation and (b) a vertical beam deviation from Trim01 (see Fig. 1) set at 0.1 A.

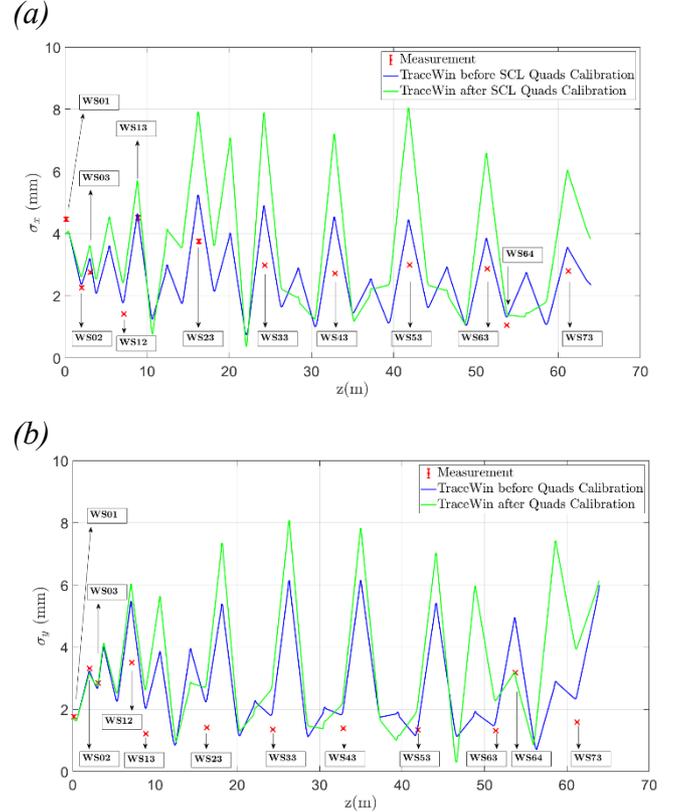

Figure 5: (a) Horizontal and (b) Vertical beam sizes measured and reconstructed with TraceWin, at all 12 Wire Scanners of the Transition and SCL sections. Simulations are performed before/after quads calibration from Linac_Gen.

The Fermilab code Linac_Gen can optimize all 32 quadrupoles of the Transition and SCL sections and find a new quadrupole calibration that matches the 29 BPM measured centroids in both planes, as shown in Figure 4. The new quadrupole calibrations from Linac_Gen stay within 10% of the initial calibration. Figure 5 shows a measured envelope along the 12 Wire Scanners with TraceWin predictions, with the initial quadrupole calibrations and the suggested ones from Linac_Gen, based on Figure 4. We can see from Figure 5 that the measured beam sizes and those from TraceWin do not agree. We think that further work is needed in the characterization of the longitudinal dynamics in the SCL section (phase and field of the 28 SCL cavities) to obtain a better agreement between the measured and expected beam size.